\documentclass[aps,preprint,onecolumn,nofootinbib]{revtex4}
\usepackage{dcolumn}
\usepackage{bm}
\usepackage{mathrsfs}
\usepackage{amssymb}
\usepackage{amsmath}
\usepackage{amsfonts}
\usepackage{color}
\usepackage{ulem}
\usepackage[]{graphicx}
\begin{document}
\bibliographystyle{prsty}
\title{Justifying Born's rule $P_\alpha=|\Psi_\alpha|^2$ using deterministic chaos, decoherence, and the de Broglie-Bohm quantum theory}
\author{Aur\'elien Drezet$^{1}$}
\address{$^1$Univ. Grenoble Alpes, CNRS, Grenoble INP, Institut Neel, F-38000 Grenoble, France}

\email{aurelien.drezet@neel.cnrs.fr}
\begin{abstract}
In this work we derive Born's rule from the pilot-wave theory of de Broglie and Bohm. Based on a toy model involving a particle coupled to a environement made of ``qubits'' (i.e., Bohmian pointers) we show that entanglement together with deterministic chaos lead to a fast relaxation from any statistitical distribution $\rho(x)$ (of finding a particle at point $x$) to the Born probability law $|\Psi(x)|^2$. Our model is discussed in the context of Boltzmann's kinetic theory and we demonstrate a kind of H theorem for the relaxation to the quantum equilibrium regime.
\end{abstract}

 \maketitle

\section{Introduction and motivations}
\label{section1}
\indent The work of Wojciech H. Zurek is universally recognized for its central importance in the field of quantum foundations in particular concerning decoherence and the understanding of the elusive border between the quantum and classical realms~\cite{Zurek1}. Zurek emphasized the role of pointer states and environment induced superselection rules (einselections). In the recent years, part of his work going beyond mere decoherence and averaging focussed on quantum darwinism and envariance. Quantum darwinism main goal is to emphasize the role of multiple copies of information records contained in the local quantum environment.  Envariance aims is to justify  the existence and form of quantum probabilities, i.e., deriving  Born's rule, from specific quantum symmetries based on entanglement~\cite{Zurek2}.  In recent important reviews of his work Zurek stressed the importance of some of theses concepts for discussing the measurement problem in relation with various interpretations of quantum mechanics~\cite{Zurek3,Zurek4}. Recent works showed for instance the importance of such envariance to establish Born's rule in the many-worlds and many-minds contexts~\cite{Wallace,Drezet2021}. While in his presentations Zurek  generally  advocated a neutral position perhaps located between the Copenhagen and Everett interpretations we believe his work on entanglement and decoherence could have a positive  impact on other interpretations such as the de Broglie Bohm theory.  We know that Zurek has always been prudent concerning Bohmian mechanics (see for example his remarks in \cite{Zurek1996} p.~209) perhaps because of the strong ontological price one has to pay in order to assume a nonlocal quantum potential and surrealistic trajectories (present even if we include decoherence~\cite{Zurek3,Appleby}). Moreover, the aim of this work is to discuss the pivotal role that quantum entanglement with an environment of ``Bohmian pointers'' could play in order to justify Born's rule in the context of such a Bohmian  interpretation. The goal is thus to suggest interesting and positive  implications that decoherence could have on ontologies different from  Everettian or Consistent histories approaches.    \\
\indent  The de Broglie-Bohm quantum theory (BBQT) introduced by de Broglie in 1927 \cite{debroglie1927,Valentini,debroglie1930} and rediscovered by Bohm in 1952 \cite{Bohm1952a,Bohm1952b}, is nowadays generally accepted as a satisfactory interpretation of quantum mechanics, at least for problems dealing with non-relativistic systems \cite{Hiley,Holland}. Within this regime the BBQT is a clean deterministic formulation of quantum mechanics preserving the classical concepts of point-like particles moving continuously in space-time. This formulation is said to be empirically equivalent to the orthodox description axiomatized by the Copenhagen school, meaning that the BBQT is able to justify and reproduce the probabilistic predictions made by the standard quantum measurement theory. More specifically, this implies recovering the famous Born rule that connects the probability \begin{eqnarray}
P_\alpha=|\Psi_\alpha|^2
\end{eqnarray}  of observing an outcome $\alpha$ (associated with the quantum  observable $\hat{A}$) to the amplitude $\Psi_\alpha $ in the quantum state expansion $|\Psi\rangle=\sum_\alpha\Psi_\alpha |\alpha\rangle$ (i.e., $|\alpha\rangle$ is an eigenstate of $\hat{A}$ for the observable eigenvalue  $\alpha$).\\
\indent This issue has been a recurrent subject of controversies since the early formulation of the BBQT (see for example  Pauli's objection in \cite{Pauli1953,Keller1953}). It mainly arizes because the BBQT is a deterministic mechanics and therefore, like for classical statistical mechanics, probabilities in the BBQT can only be introduced in relation with ignorance and uncertainty on the initial conditions of the particle motions. Moreover, after more than one and a half century of developments since Maxwell and Boltzmann times, it is well recognized that the physical and rigorous mathematical foundation of statistical mechanics is still debatable \cite{Uffink}. The BBQT, which in some sense generalizes and extends Newtonian mechanics, clearly inherits of these difficulties constituting strong obstacles for defining a clean basis of its statistical formulation. This fact strongly contrasts with standard quantum mechanics for which randomness has been from the start axiomatized as genuine and inevitable.\\
\indent Over the years several responses have been proposed by different proponents of the BBQT to justify Born's rule (for recent reviews see \cite{Barrett1995,Callender2007,Drezet2017}). Here, we would like to focus on the oldest approach which goes back to the work of David Bohm on deterministic and molecular chaos. Indeed, already in 1951-1952 Bohm emphasized the fundamental role of the disorder and chaotic motion of particles for justifying Born's rule \cite{Bohm1952a,Bohm1952b}. In his early work Bohm stressed that the complexity of the de Broglie-Bohm dynamics during interaction processes, such as quantum measurements, should drive the system to quantum equilibrium. In other words, during interactions with an environment like a measurement apparatus any initial probability distribution $\rho(X)\neq |\Psi(X)|^2$ for particle in the configuration space (here $X=[\mathbf{x}_1,...,\mathbf{x}_M]\in \mathbb{R}^{3M}$ is a vector in the $N$-particles configuration space) should evolve in time to reach the quantum equilibrium limit $\rho(X)\rightarrow|\Psi(X)|^2$ corresponding to Born's rule. In this approach the relaxation process  would be induced by both the high sensitivity to changes in the initial conditions of the particle motions (one typical signature of deterministic chaos) and by the molecular thermal chaos resulting from the macroscopic nature of the interacting environment (i.e., with $\sim 10^{23}$ degrees of freedom). Furthermore, in this strategy Born's rule $\rho(X)=|\Psi(X)|^2$ should appear as an attractor similar to the microcanonical and canonical ensemble in thermodynamics. In 1953 Bohm developed an example model \cite{Bohm1953} (see \cite{Potel2002} for a recent investigation of this idea) where a quantum system randomly submitted to several collisions with external particles constituting a bath was driven to quantum equilibrium  $\rho(X)=|\Psi(X)|^2$. In particular, during his analysis Bohm sketched a quantum version of the famous Boltzmann $H$-theorem for proving the irreversible tendency to reach Born's rule (for other clues that Bohm was strongly fascinated by deterministic chaos already in the 1950's see \cite{Bohm1955} and the original 1951 manuscript written by Bohm in 1951~\cite{Bohm1951} and rediscovered recently). However, in later works, specially in the one made with Vigier \cite{BohmVigier1954} and then subsequently Hiley \cite{Hiley}, Bohm modified the original de Broglie-Bohm dynamics by introducing stochastic and fluctuating elements associated with a subquantum medium forcing the relaxation towards quantum equilibrium $\rho(X)\rightarrow|\Psi(X)|^2$.\\
\indent While this second semi-stochastic approach was motivated by general philosophical considerations \cite{Bohm1957} proponents of the BBQT have felt divided concerning the need for such a modification of the original framework. In particular, starting in the 1990's Valentini currently develops an approach assuming the strict validity of the BBQT as an underlying deterministic framework and introduces mixing and coarse-graining \`a la Tolman-Gibbs in the configuration space in order to derive a Bohmian `subquantum' version of the $H$-theorem \cite{Valentini1991,ValentiniPhD}. Importantly, in his work on the `subquantum heat-death' (i.e., illustrated with many numerical calculations \cite{Valentini2005,Valentini2012} often connected with cosmological studies \cite{Valentini2007,Valentini2015}) Valentini and coworkers stress the central role of deterministic chaos in the mixing processes. Moreover, deterministic chaos in the BBQT is a research topic in itself (for a recent review see \cite{Efthymiopoulos2017,Contopoulos2020}) and many authors  (including Bohm \cite{Hiley} and Valentini \cite{Valentini2005,Valentini2012}) have stressed the role of nodal-lines associated with phase-singularities of the wavefunction for steering deterministic chaos in the BBQT \cite{Frisk1997,Falsaperia2003,Wisniacki2005}. Yet, it has also been pointed out \cite{Efthymiopoulos2017,Efthymiopoulos2012} that such a chaos is not generic enough to force  the quantum relaxation $\rho(X)=|\Psi(X)|^2$ for any arbitrary initial conditions $\rho(X)\neq |\Psi(X)|^2$ (a reversibility objection \`a la Kelvin-Loschmidt is already sufficient to see the impossibility of such an hypothetical deduction \cite{Callender2007,Norsen2018}).\\
\indent In the present work we emphasize the role of an additional ingredient which together with chaos and coarse graining can help and steer the quantum dynamical relaxation $\rho(X)\rightarrow|\Psi(X)|^2$, namely, quantum entanglement with the environment. The idea that quantum correlations must play a central role in the BBQT for justifying Born's rule is not new of course. Bohm already emphasized the role of entanglement in his work \cite{Bohm1952b,Hiley,Bohm1953}. Moreover, in recent studies motivated by the Vigier-Bohm analysis we developed a Fokker-Planck \cite{Drezet2017} and Langevin's like \cite{Drezet2018} description of relaxation to quantum equilibrium $\rho(X)=|\Psi(X)|^2$ by coupling a small system $S$ to a thermal bath or reservoir $T$ inducing a Brownian motion on $S$. We showed that under reasonable assumptions we can justify a version of the $H$-theorem where quantum equilibrium appears as a natural attractor. Furthermore, at the end of \cite{Drezet2017}  we sketched a even simpler strategy based on mixing together with entanglement and involving deterministic chaotic iterative maps. After the development of such an idea it came to our attention that a similar strategy has been already developed in a elegant work by Philbin \cite{Philbin2015} and therefore we didn't include too much details concerning our model in \cite{Drezet2017}. Here, we give the missing part and provide a more complete and quantitative description of our scenario presented as an illustration of a more general scheme. More precisely, we  i) analyze the chaotic character of the specific de Broglie Bohm dynamics associated with our toy model, ii) build a Boltzmann diffusion equation for the probability distribution, and finally  iii) derive a simple $H$-theorem from which Born's rule turns out to be an attractor.  We emphasize that our work, like the one of Philbin, suggests interesting future developments for justifying Born's rule and recovering standard quantum mechanics within the BBQT.   
\section{The status of Born's rule in the de Broglie Bohm theory}
\label{section2}
\indent We start with the wavefunction $\psi(\mathbf{x},t)=R(\mathbf{x},t)e^{iS(\mathbf{x},t)/\hbar}$ obeying to Schr\"odinger's equation 
\begin{eqnarray}
i\hbar \frac{\partial}{\partial t}\psi(\mathbf{x},t)=\frac{-\hbar^2\boldsymbol{\nabla}^2}{2m}\psi(\mathbf{x},t)+V(\mathbf{x},t)\psi(\mathbf{x},t)\label{eq1} 
\end{eqnarray}
for a single nonrelativistic particle with masses $m$ in the external potentials $V(\mathbf{x},t)$ (we limit the analysis to a single particle but the situation is actually generic).  The BBQT leads to the first-order `guidance' law of motion
\begin{eqnarray}
\frac{d}{dt}\mathbf{x}^\psi(t)=\mathbf{v}^\psi(\mathbf{x}^\psi(t),t)\label{eq2} 
\end{eqnarray}  where $\mathbf{v}^\psi(\mathbf{x},t)=\frac{1}{m}\boldsymbol{\nabla}S(\mathbf{x},t)$ defines an Eulerian velocity field and $\mathbf{x}^\psi(t)$ is a de Broglie Bohm particle trajectory. Furthermore, from Eq. \ref{eq1} we obtain the conservation rule:
\begin{eqnarray}
-\frac{\partial}{\partial t}R^2(\mathbf{x},t)=\boldsymbol{\nabla}\cdot[R^2(\mathbf{x},t)\mathbf{v}^\psi(\mathbf{x},t)]
\label{eq3} 
\end{eqnarray} where we recognize $R^2(\mathbf{x},t)=|\psi(\mathbf{x},t)|^2$ the distribution which is usually interpreted as Born's probability density. Now, in the abstract probability theory we assign to every points $\mathbf{x}$ a density $\rho(\mathbf{x},t)$ corresponding to a fictitious conservative fluid obeying the constraint
\begin{eqnarray}
-\frac{\partial}{\partial t}\rho(\mathbf{x},t)=\boldsymbol{\nabla}\cdot[\rho(\mathbf{x},t)\mathbf{v}^\psi(\mathbf{x},t)].
\label{eq3b} 
\end{eqnarray}Comparing with Eq.~\ref{eq3} we deduce that the normalized distribution $f(\mathbf{x},t)=\frac{\rho(\mathbf{x},t)}{R^2(\mathbf{x},t)}$ satisfies the equation 
\begin{eqnarray}
[\frac{\partial}{\partial t}+\mathbf{v}^\psi(\mathbf{x},t)\cdot\boldsymbol{\nabla}]f(\mathbf{x},t):=\frac{d}{dt}f(\mathbf{x},t)=0.
\label{eq3c} 
\end{eqnarray} This actually means \cite{Bohm1953}  that $f$ is an integral of motion along any trajectory $\mathbf{x}^\psi(t)$. In particular, if $f(\mathbf{x},t_{in})=1$ at a given time $t_{in}$ and for any point $\mathbf{x}$ this holds true at any time $t$. Therefore, Born's rule being valid at a given time will be preserved at any other time \cite{debroglie1930,Bohm1952a,Bohm1953}. At that stage the definition of the probability  $\rho(\mathbf{x},t)d^3\mathbf{x}$ for finding a particle in the infinitesimal volume $d^3\mathbf{x}$  is rather formal and corresponds to a Bayesian-Laplacian interpretation where probabilities are introduced as a kind of measure of chance.     \\     
\indent Moreover, in BBQT the actual and measurable density of particles must be defined using a collective or ensemble of $N$ independent systems prepared in similar quantum states $\psi(\mathbf{x}_i,t)$ with $i=1,...,N$.  Yet, the concept of independency in quantum mechanics imposes the whole statistical ensemble with $N$ particles to be described by the total factorized wavefunction   
\begin{eqnarray}
\Psi_N(\mathbf{x}_1,...,\mathbf{x}_N,t)=\prod_{i=1}^{i=N}\psi(\mathbf{x}_i,t)\label{eq4} 
\end{eqnarray}
solution of the equation \begin{eqnarray}
i\hbar \frac{\partial}{\partial t}\Psi_N=[\sum_{i=1}^{i=N}\frac{-\hbar^2\boldsymbol{\nabla}_i^2}{2m}+V(\mathbf{x}_i,t)]\Psi_N.
\end{eqnarray}
For this quantum state $\Psi_N$ the BBQT allows us to build the velocity fields $
\frac{d}{dt}\mathbf{x}^\psi_i(t)=\mathbf{v}^\psi(\mathbf{x}^\psi_i(t),t)$ where $\mathbf{x}^\psi_i(t):=\mathbf{x}^{\Psi_N}_i(t)$ define the de Broglie Bohm paths for the uncorrelated particles (i.e., guided by the individual and independent wave functions $\psi(\mathbf{x}_i,t)$ and Eulerian flows $\mathbf{v}^{\Psi_N}_i(\mathbf{x}_1,...,\mathbf{x}_N,t)=\mathbf{v}^\psi(\mathbf{x}_i,t)$). Within this framework the actual density of particles  $P(\mathbf{r},t)$ at point $\mathbf{r}$
is given by 
\begin{eqnarray}
P(\mathbf{r},t)=\frac{1}{N}\sum_{k=1}^{k=N}\delta^3(\mathbf{r}-\mathbf{x}^{\psi}_k(t))
\label{eq5} 
\end{eqnarray} which clearly obeys the conservation rule
\begin{eqnarray}
-\frac{\partial}{\partial t}P(\mathbf{x},t)=\boldsymbol{\nabla}\cdot[P(\mathbf{x},t)\mathbf{v}^\psi(\mathbf{x},t)].
\label{eq6} 
\end{eqnarray}
Comparing with Eq. \ref{eq3c} we see that 
if $\rho(\mathbf{x},t)=f(\mathbf{x},t)|\psi(\mathbf{x},t)|^2$ plays the role of an abstract Laplacian probability, $P(\mathbf{r},t)$ instead represents the frequentist statistical probability. Both concepts are connected by the weak law of large numbers (WLLN) which is demonstrated in the limit $N\rightarrow +\infty$ and leads to the equality $\rho(\mathbf{x},t)=P(\mathbf{r},t)$, i.e., 
  \begin{eqnarray}
f(\mathbf{r},t)|\psi(\mathbf{r},t)|^2=\equiv \lim_{N \to +\infty}\frac{1}{N}\sum_{k=1}^{k=N}\delta^3(\mathbf{r}-\mathbf{x}^{\psi}_k(t))
\label{eq5b} 
\end{eqnarray} where the equality must be understood  in the sense of  a `limit in probability' based on typicality and not as the more usual `point-wise limit'. We stress that the application of the WLLN already relies on the Laplacian notion of measure of chance since by definition in a multinomial Bernoulli process the abstract probability density $\rho_N(\mathbf{x}_1,...,\mathbf{x}_N,t)=\prod_{i=1}^{i=N}\rho(\mathbf{x}_i,t)$ is used for weighting an infinitesimal volume of the $N$-particles configuration space $d\tau_N:=\prod_{i=1}^{i=N}d^3\mathbf{x}_i$. It can be shown that in the limit $N\rightarrow +\infty$ with the use of this measure $\rho_Nd\tau_N$ almost all possible configurations $\mathbf{x}^{\psi}_1(t),...,\mathbf{x}^{\psi}_N(t)$ are obeying the generalized Born's rule $P(\mathbf{r},t)=\rho(\mathbf{x},t)=f(\mathbf{x},t)|\psi(\mathbf{x},t)|^2$ (the fluctuation varying as $\frac{1}{\sqrt{N}}$). It is in that sense that Eq. \ref{eq5b} is said to be typical where typical meaning valid for `overwhelmingly many', i.e., almost all configurations in the whole configuration space weighted by $\rho_Nd\tau_N$. The application of the law of large number to the BBQT is well known and well established \cite{ValentiniPhD,Durr1992a,Durr1992b} but has been the subject of intense controversies \cite{Norsen2018,Durr1992a,Valentini2020,Durr2019}. Issues concern 1) the interpretation of $\rho_N$ as a probability density, i.e., in relation with the notion of typicality, and 2) the choice of $f=1$ as natural and guided by the notion of equivariance \cite{remark}. To paraphrase David Wallace the only thing the law of large numbers proves is that relative frequency tends to weight ... with high weight \cite{Wallacevideo}. However, there is a certain circularity in the reasoning which at least shows that the axiomatic of the probability calculus allows us to identify an abstract probability such as $\rho d^3\mathbf{x}$ to a frequency of occurrence like $P d^3\mathbf{x}$. However, the law is unable to guide us for selecting the good measure for weighting typical configurations (the condition on equivariance \cite{remark} is only a mathematical recipe not a physical consequence of a fundamental principle). Therefore, the value of the $f$ function is unconstrained by the typicality reasoning without already assuming the result \cite{Valentini2020}. In other words, it is impossible to deduce Born's rule without already assuming it.\\
\indent  Moreover, it is important to see that the relation $\frac{d}{dt}f(\mathbf{x}^\psi(t),t)=0$ is playing the same role in the BBQT for motions in the configuration space that Liouville's theorem $\frac{d}{dt}\eta(q(t),p(t),t)=0$ in classical statistical mechanics  (where $\eta(q,p,t)$ is the probability density in phase space $q,p$). Therefore, with respect to the measure $d\Gamma=|\psi(\mathbf{x},t)|^2d^3\mathbf{x}$ (which is preserved in time along trajectories since $\frac{d}{dt}d\Gamma_t=0$), the condition $f=1$ is equivalent to the postulate of equiprobability used in standard statistical mechanics for the microcanonical ensemble. Clearly, we see that the inherent difficulties existing in classical statistical mechanics for justifying the microcanonical ensemble are transposed in the  BBQT for justifying Born's rule, i.e., $f=1$.       
\section{A deterministic and chaotic model for recovering Born's rule within the de Broglie Bohm quantum theory}\label{section3}
\subsection{The basic dynamics} \label{section31}
 \indent As a consequence of the previous discussion we now propose a simple toy model where the condition $f=1$ appears as an attractor, i.e, $f_t \rightarrow 1$ during a mixing process.
 We consider a single electron wave-packet impinging on a beam-splitter. To simplify the discussion we consider an incident wave-train with one spatial dimensions $x$  characterized by the wavefunction 
 \begin{eqnarray}
    \psi_0(x,t)\simeq \Phi_0(x-v_x t)e^{i(k_x x-\omega_k t)} \label{num0} 
 \end{eqnarray} 
 where we have the dispersion relation $E_k:=\hbar\omega_k=\frac{\hbar^2k_x^2}{2m}$ and the (negative) group velocity components $v_x=\frac{\hbar k_x}{m}<0$ with $k_x=-|k_x|$. Furthermore, for mathematical consistency we impose $\Phi_0\simeq const.=C$ in the spatial support region where the wave-packet is not vanishing and the typical wavelength $\lambda=2\pi/|k_x| \ll L$ where $L$ is a typical wave-packet spatial extension. If we assume Born's rule $|C|^2$ must be identified with a probability density  and by normalization this implies $C=1/\sqrt{L}$ (this point will be relevant only in Sec.~\ref{section3}). The beam-splitter is a rectangular potential barrier or well $V(x)=V_0$ with $V_0$ a constant  in the  region $|x|<\epsilon/2\ll L$ and $V(x)=0$ otherwise.
 During the interaction with the beam-splitter the whole wavefunction approximately reads
\begin{eqnarray}
\psi(x,t)\simeq \psi_0(x,t)+ R_k \psi_1(x,t) \nonumber \\\textrm{if $x>\epsilon/2$ }\nonumber \\
     \psi(x,t)\simeq \Phi_0(-v_x t)[A_ke^{iq_x x}+B_k e^{-iq_x x}]e^{-i\omega_k t}\nonumber \\  \textrm{ if $|x|<\epsilon/2$ }\nonumber \\
    \psi(x,t)\simeq T_k\psi_0(x,t)\nonumber \\ \textrm{ if $x<-\epsilon/2$ }
\end{eqnarray} 
where $\psi_1(x,t)=\Phi_0(x+v_x t)e^{-ik_x x}e^{-i\omega_k t}=\psi_0(-x,t)$, and $R_k$ (reflection amplitude), $T_k$(transmission amplitude) and $A_k,B_k$ are Fabry-Perot coefficients computed in the limit where the wave-packet is infinitely spatially extended. 
 We have: 
 \begin{eqnarray}
T_k=\frac{4q_xk_x}{(q_x+k_x)^2}\frac{1}{e^{i(q_x-k_x)\epsilon}-\frac{(q_x-k_x)^2}{(q_x+k_x)^2}e^{-i(k_x+q_x)\epsilon}}\nonumber \\
R_k=iT_k\frac{k_x^2-q_x^2}{2q_xk_x}\sin{(q_x\epsilon)}\nonumber \\
A_k=R_k[\frac{q_x+k_x}{2q_x}e^{-i(q_x-k_x)\epsilon/2}+\frac{q_x-k_x}{2q_x}e^{-i(q_x+k_x)\epsilon/2}]\nonumber \\
B_k=R_k[\frac{q_x-k_x}{2q_x}e^{i(q_x+k_x)\epsilon/2}+\frac{q_x+k_x}{2q_x}e^{i(q_x-k_x)\epsilon/2}]\nonumber \\
 \end{eqnarray} where $q_x$ is given by the dispersion relation $E_k:=\hbar\omega_k=\frac{\hbar^2q_x^2}{2m}+V_0$, i.e., $q_x^2-k_x^2=-2mV_0/\hbar$. As an 
 illustration we choose $\epsilon=\frac{1}{2}\frac{\lambda}{2\pi}$ and $q_x\simeq 2.5 k_x$  (i.e., $V_0<0$) which leads to $T_k\simeq \frac{1}{\sqrt{2}}e^{i0.267\pi}$ and $R_k=iT_k$ corresponding  to a balanced 50/50 beam-splitter.\\  
\indent We need now to consider the problem from the point of view of the scattering matrix theory. First, observe that for negative time $t_{in}<0$  (with $|t_{in}|\gg L/|v_x|$) the incident wave-packet $\psi_0(x,t_{in})$ given by Eq. \ref{num0} which is coming from the $x>0$ region  with a negative group velocity is transformed for large positive times $t_f>0$ (with $|t_f|\gg L/|v_x|$) into the two non overlapping wave-packets 
\begin{eqnarray}
    \psi(x,t_f)\simeq  R_k\psi_1(x,t_f) & & \textrm{if $x>0$ }\nonumber \\
    \psi(x,t_f)\simeq T_k\psi_0(x,t_f) &&\textrm{if $x<0$ }.
 \end{eqnarray} Since the wave packets are non-overlapping we write: 
  \begin{eqnarray}
  \psi(x,t_f)\simeq R_k\psi_1(x,t_f)+T_k\psi_0(x,t_f). 
 \end{eqnarray}
Of course, the situation is symmetric: if an incident wave-packet $\psi_1(x,t_{in})$  comes from the $x<0$ region with a positive group velocity for $t_{in}<0$ we will obtain at the end, i.e., for $t_f>0$:
\begin{eqnarray}
  \psi(x,t_f)\simeq T_k\psi_1(x,t_f)+R_k\psi_0(x,t_f). 
 \end{eqnarray}
The general case can thus be treated by superposition: an arbitrary initial state $\psi(x,t_{in})=a_+\psi_0(x,t_{in})+a_-\psi_1(x,t_{in})$ for negative times $t_{in}$ (with $|t_{in}|\gg L/|v_x|$) will evolve into
  \begin{eqnarray}
  \psi(x,t_f)\simeq (a_+R_k +a_-T_k)\psi_1(x,t_f) \nonumber \\ + (a_+T_k +a_-R_k)\psi_0(x,t_f) \label{sum}
 \end{eqnarray} for positive times $t_f$ (with $|t_f|\gg L/|v_x|$). Writing $a'_+=a_+R_k +a_-T_k$ and $a'_-=a_+T_k +a_-R_k$ the different mode amplitudes we define a 2$\times$2 unitary transformation
 \begin{eqnarray}
  \left(\begin{array}{c}
  a'_+\\a'_-
  \end{array}\right)=\left(\begin{array}{cc}
  R_k &T_k\\ T_k & R_k
  \end{array}\right)\left(\begin{array}{c}
  a_+\\a_-
  \end{array}\right)\nonumber \\
  =\frac{e^{i0.267\pi}}{\sqrt{2}}\left(\begin{array}{cc}
  i &1\\ 1 & i
  \end{array}\right)\left(\begin{array}{c}
  a_+\\a_-
  \end{array}\right).\label{arr}
 \end{eqnarray}
 \indent Moreover, consider now the point of view of the BBQT. Following this theory the dynamics of the material point is obtained by integration of the guidance equation
  \begin{eqnarray}
\frac{d}{dt}x^\psi(t)=v^\psi(x^\psi(t),t)=\frac{\hbar}{m}\textrm{Im}[\frac{\partial}{\partial x }\psi(x,t)|_{x=x^\psi(t)} ]\label{Neq2} 
\end{eqnarray} 
\begin{figure}[h]
\begin{center}
\includegraphics[width=10cm]{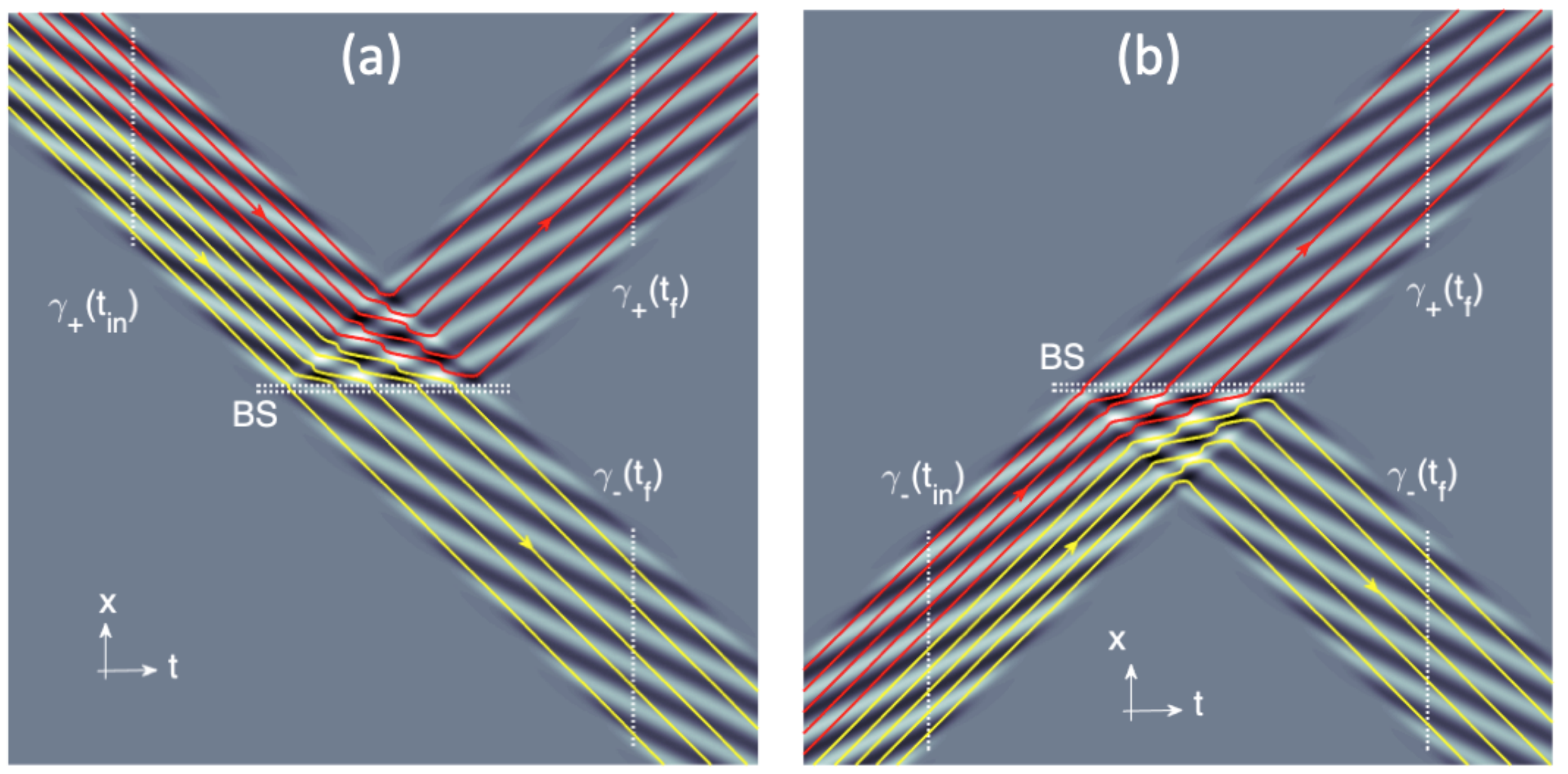} 
\caption{(a) Scattering of a 1D wavepacket impinging  on a  50/50 beam-splitter (BS). The  colormap shows $Re[\Psi(x,t)]$ in the $t,x$ plane. The color (red and  yellow) lines are de Broglie-Bohm trajectories associated with this wave-function (red  and yellow trajectories are ending in two different wave-packets. The dotted white lines are crosscuts as discussed in the main text.  (b) Shows a similar situation when a wave-packet impinges on the other input gate.  } \label{Fig1}
\end{center}
\end{figure}
which can easily be computed numerically. We illustrate in Fig. \ref{Fig1} the interaction with the 50/50 beam-splitter  characterized by Eq. \ref{arr}  of a rectangular wave-packet  (i. e., $\Phi_0(x)=C$ if $|x|<L/2$  where $L$ is the width of the wave-packet) incident from the $x>0$ region (i.e., $a_+=1,a_-=0$). As a remarkable feature we can see the so called Wiener fringes \cite{debroglie1930} existing in the vicinity of the beam-splitter and which strongly alter the de Broglie-Bohm trajectories.  What is also immediately visible is that the de Broglie Bohm trajectories $x^\psi(t)$ never cross each other. This is a general property of this first order dynamics \cite{Holland,Hiley} which plays a central role in our analysis.\\ \indent An interesting feature of this example concerns the density of `probability' $|\psi(x,t)|^2$. Indeed, consider a time $t_{in}$ in the remote past before the wave-packet coming from the positive region (i.e. like in Fig. \ref{Fig1}) interacts with the potential well. At that time the center of the wave-packet is located at $x_{in}=v_xt_{in}>0$. Yet, since trajectories can not cross each others we know that the ensemble $\gamma_+(t_{in})$ of all possible particle positions at time $t_{in}$, i.e., $x^\psi(t_{in})\in [x_{in}-\frac{L}{2},x_{in}+\frac{L}{2}]$ is divided into two parts. In the first part  
$\gamma_+^{(+)}(t_{in})$, i.e., $x^\psi(t_{in})\in [x_{in}+H,x_{in}+\frac{L}{2}] $ with $|H|<\frac{L}{2}$, all particles evolve in the future  (i.e., at time $t_f$) into the $\psi_1(x,t_f)$ reflected  wave-packet  (corresponding to the support $\gamma_+(t_f)$, i.e., $x^\psi(t_f)\in [x_f-\frac{L}{2},x_f-\frac{L}{2}]$ with $x_f=-v_xt_f>0$). In the second part  $\gamma_+^{(-)}(t_{in})$, i.e., $x^\psi(t_{in})\in [x_{in}-\frac{L}{2},x_{in}+H]$, all the particles necessarily end up their journey in the $\psi_0(x,t_f)$  transmitted wave-packet (corresponding to the support $\gamma_-(t_f)$, i.e., $x^\psi(t_f)\in [-x_f-\frac{L}{2},-x_f-\frac{L}{2}]$). Now, remember that from the de Broglie-Bohm Liouville theorem the measure $d\Gamma(x,t)=|\psi(x,t)|^2dx$ is preserved in time, i.e., $\frac{d}{dt}d\Gamma_t=0$. Therefore,  the measure \begin{eqnarray}
\Gamma_+(t_f)=\int_{\gamma_+}|\psi(x,t)|^2dx=LC^2/2
\end{eqnarray} associated with the reflected wave  necessarily equals the measure associated with the segment  $\gamma_+^{(+)}(t_{in})$, i.e., 
\begin{eqnarray}
\Gamma_+^{(+)}(t_{in})=(L/2-H)C^2=\Gamma_+(t_f).
\end{eqnarray}  This leads to $H=0$ which in turn means that  $\gamma_+^{(+)}(t_{in})$ corresponds to $x^\psi(t_{in})\in [x_{in},x_{in}+\frac{L}{2}]$ and $\gamma_+^{(-)}(t_{in})$ to $x^\psi(t_{in})\in [x_{in}-\frac{L}{2},x_{in}]$. This result is actually general  and holds for any symmetric wave-packet  $\Phi_0(x)=\Phi_0(-x)$ if we can neglect the overlap between $\Phi_0(x-v_x t_f)$ and $\Phi_0(x+v_x t_f)$).\\ 
\indent Moreover,
\begin{figure}[h]
\begin{center}
\includegraphics[width=10cm]{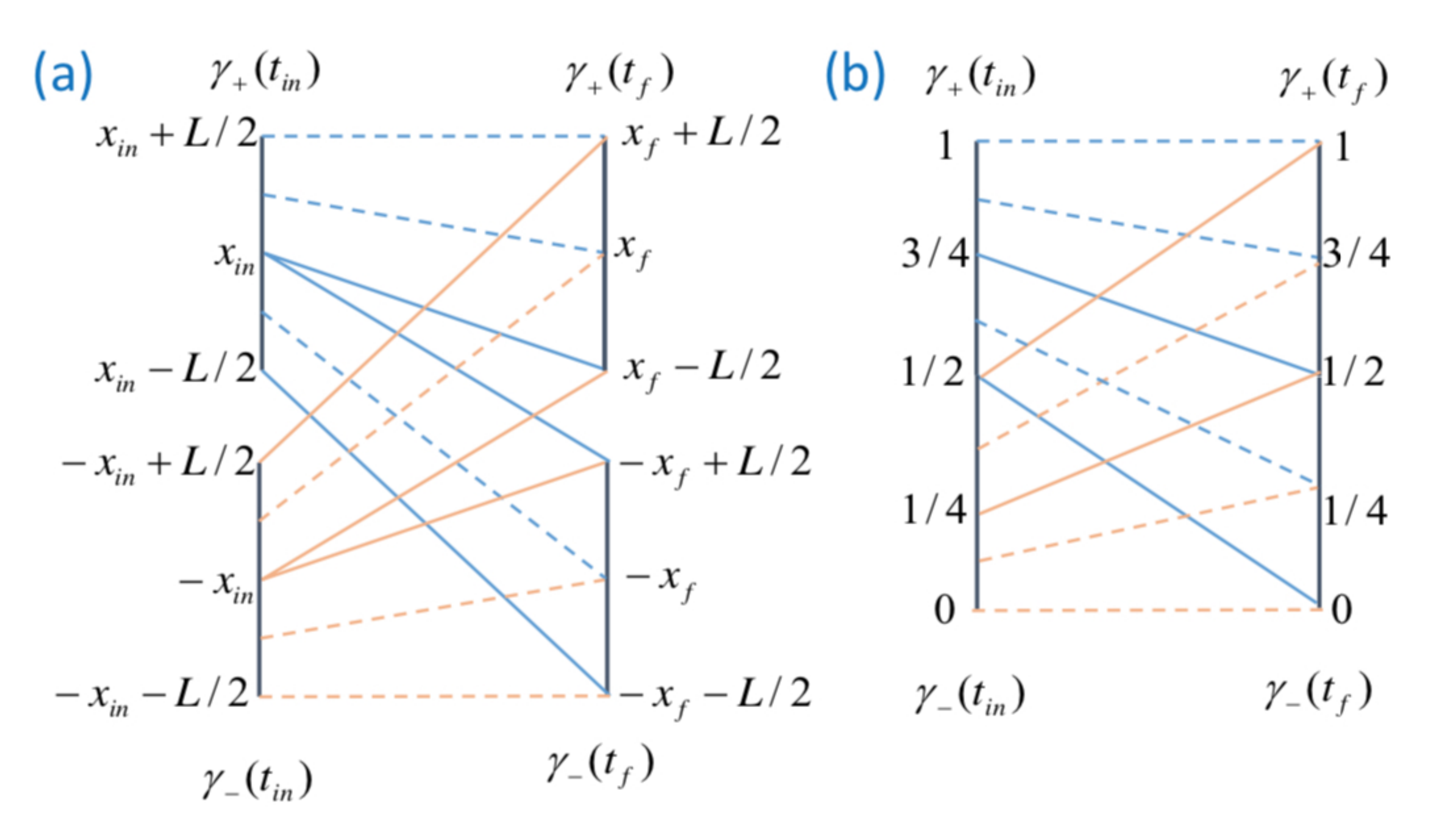} 
\caption{(a) Transformation from the initial $\gamma_{\pm}(t_{in})$ $x-$space to the final $\gamma_{\pm}(t_{ou})$ $x-$space for the two situations shown Fig.~\ref{Fig1}(a) and Fig.~\ref{Fig1}(b) respectively (depicted as blue lines and red lines respectively). (b) Shows the same transformation using the $y$ coordinate instead of the $x$ coordinate (as explained in the main text).} \label{Fig2}
\end{center}
\end{figure}
for the rectangular wave-packet we deduce from the de Broglie-Bohm Liouvillian theorem $\frac{d}{dt}d\Gamma_t=0$ that any infinitesimal length element $\delta x^\psi(t_{in})$  surrounding a point  $x^\psi(t_{in})$ in $\gamma_+(t_{in})$ evolves to the infinitesimal length  $\delta x^\psi(t_f)=2\delta x^\psi(t_{in})$ surrounding the point $x^\psi(t_f)$ located in $\gamma_\pm(t_f)$. This property can be used to define a simple mapping between the initial coordinates $x^\psi(t_{in})\in \gamma_+(t_{in})$ and the final outcome $x^\psi(t_f)\in \gamma_+(t_f)\cup\gamma_-(t_f)$. It is simpler to introduce the normalized variables: 
\begin{eqnarray}
y(t_{in})=\frac{x^\psi(t_{in})-x_{in}}{2L}+\frac{3}{4}\in [\frac{1}{2},1] &  & \textrm{  if $x^\psi(t_{in})\in \gamma_+(t_{in})$ }\nonumber\\
y(t_f)=\frac{x^\psi(t_f)-x_f}{2L}+\frac{3}{4}\in [\frac{1}{2},1] &  & \textrm{  if $x^\psi(t_f)\in \gamma_+(t_f)$ }\nonumber  \\
y(t_f)=\frac{x^\psi(t_f)+x_f}{2L}+\frac{1}{4}\in [0,\frac{1}{2}] &  & \textrm{  if $x^\psi(t_f)\in \gamma_-(t_f)$ }.\nonumber  \\ \label{transf1}
\end{eqnarray}
 The mapping between the two new ensembles  (that we will continue to name $\gamma_+(t_{in})$ and $\gamma_+(t_f)\cup\gamma_-(t_f)$) is thus simply written:  
 \begin{eqnarray}
y(t_f)=2y(t_{in})-1.\label{mod1}
\end{eqnarray} The result of this mapping is illustrated using the $x$ coordinates in Fig. \ref{Fig2}(a) or the $y$ coordinates in Fig. \ref{Fig2}(b). In particular, it is visible that the correspondence $y(t_f)=F(y(t_{in}))$ is not always univocally defined.  This occurs at $x^\psi(t_{in})=x_{in}$ (i.e., $y(t_{in})=\frac{3}{4}$) which evolves either as $x^\psi(t_f)=x_f-L/2\in \gamma_+(t_f)$ or $x^\psi(t_f)=-x_f+L/2\in \gamma_-(t_f)$ corresponding to the single value $y(t_f)=\frac{1}{2}$.  Physically, as shown in Fig \ref{Fig2}(a), it means that a point located at the center of the wave-packet  $\psi_0(x,t_{in})$ is unable to decide whether it should move into the reflected or transmitted wave-packets: this is a point of instability. This apparently violates the univocity of the de Broglie-Bohm dynamics Eq. \ref{Neq2} which imposes that at given point, i.e., $x^\psi(t_{in})=x_{in}$, one and only one trajectory is defined. However, we stress that this pathology is actually a consequence of the oversimplification of our model consisting in assuming a idealized rectangular wave packet $\Phi_0(x)=C$ if $|x|<L/2$ with abrupt boundaries at $|x|=L/2$. In a real experiment with a Gaussian wave-packet the point $x^\psi(t_{in})=x_{in}$ would be mapped at the internal periphery of the two wave-packets constituting $\psi(x,t_f)$ (this would correspond to the points $x^\psi(t_f)=\pm \epsilon/2\sim 0$ where the beam splitter is located). In this regime our assumption of a finite support for $\Phi_0(x)$ is not acceptable anymore. \\ 
\indent The previous analysis was limited to the case of the wave-packet  $\psi_0(x,t_{in})$ coming from the $x>0$ region. However, in the symmetric case of a wave-packet $\psi_1(x,t_{in})$ coming from the $x<0$ region (i.e., $a_+=0,a_-=1$) the situation is very similar (as shown in Fig. \ref{Fig2}) with the only differences that the $\gamma_+(t_{in})$ space is changed into $\gamma_-(t_{in})$, i.e., $x^\psi(t_{in})\in [-x_{in}-\frac{L}{2},-x_{in}+\frac{L}{2}] $ and the roles of $\gamma_+(t_f)$ and  $\gamma_-(t_f)$ (which definitions are let unchanged) are now permuted (i.e., $\gamma_+(t_f)$ is now associated with the transmitted wave-packet and  $\gamma_-(t_f)$ with the reflected one). From the point of view of the BBQT the trajectories of Fig. \ref{Fig1}(b) are obtained by a mirror symmetry $x\rightarrow -x$ from Fig. \ref{Fig1}(a). The new mapping $ x^\psi(t_{in})\rightarrow x^\psi(t_f)$ is now well described by the variable transformation:
\begin{eqnarray}
y(t_{in})=\frac{x^\psi(t_{in})+x_{in}}{2L}+\frac{1}{4}\in [0,\frac{1}{2}] &  & \textrm{  if $x^\psi(t_{in})\in \gamma_-(t_{in})$ }\nonumber\\
y(t_f)=\frac{x^\psi(t_f)-x_f}{2L}+\frac{3}{4}\in [\frac{1}{2},1] &  & \textrm{  if $x^\psi(t_f)\in \gamma_+(t_f)$ }\nonumber  \\
y(t_f)=\frac{x^\psi(t_f)+x_f}{2L}+\frac{1}{4}\in [0,\frac{1}{2}] &  & \textrm{  if $x^\psi(t_f)\in \gamma_-(t_f)$ }.\nonumber  \\ \label{transf2}
\end{eqnarray} which lets the definition of $y(t_f)$ unchanged with respect to Eq. \ref{transf1}.  The mapping between the two ensembles $\gamma_-(t_{in})$ and $\gamma_+(t_f)\cup\gamma_-(t_f)$ is now written:  
 \begin{eqnarray}
y(t_f)=2y(t_{in})\label{mod2}
\end{eqnarray} which  is very similar to Eq. \ref{mod1}.
\subsection{Entanglement and Bernoulli's shift}\label{section32}
\indent  If we regroup Eq. \ref{transf1} and Eq. \ref{transf2} together with  Eq. \ref{mod1} and Eq. \ref{mod2} we are tempted to recognize the well known Bernoulli map:   
\begin{eqnarray}
y(t_f)=2y(t_{in})&  \textrm{mod}(1),  
 \label{mod3}
\end{eqnarray} which actually means
\begin{eqnarray}
y(t_f)=2y(t_{in})-1 & \textrm{  if $y(t_{in})>\frac{1}{2}$ } \nonumber \\
y(t_f)=2y(t_{in}) & \textrm{  if $y(t_{in})<\frac{1}{2}$ }
\label{mod3b}
\end{eqnarray} 
 for $y(t_f)$ and $y(t_{in}) \in [0,1]$. This would physically corresponds to a mapping  $\gamma_+(t_{in})\cup\gamma_-(t_{in})\rightarrow \gamma_+(t_f)\cup\gamma_-(t_f)$. In classical physics such a mapping would be unproblematic since the two dynamics given by Eq. \ref{mod1} and Eq. \ref{mod2} could be superposed without interference. However, in quantum mechanics, and specially in the BBQT, the dynamics is contextually guided by the whole wave function $\psi(x,t)$ and a general  superposition of states $\psi(x,t_{in})=a_+\psi_0(x,t_{in})+a_-\psi_1(x,t_{in})$ evolves at $t_f$ to the state $\psi(x,t_{in})$ given by Eq. \ref{sum}. Consider for example with Eq. \ref{arr} the unitary evolution
 \begin{eqnarray}
\frac{\psi_0(x,t_{in})+i\psi_1(x,t_{in})}{\sqrt{2}}\rightarrow ie^{i0.267\pi}\psi_1(x,t_f).\label{sum2}
 \end{eqnarray}
\begin{figure}[h]
\begin{center}
\includegraphics[width=6.5cm]{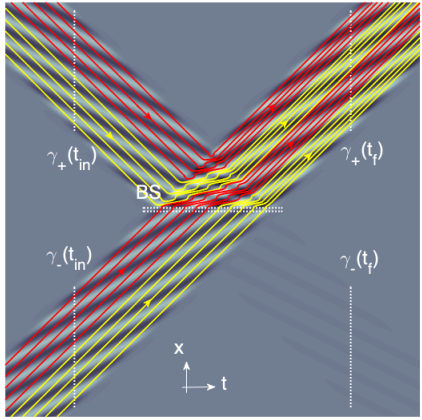} 
\caption{Same as in Fig.~\ref{Fig1} but for a symmetric superposition of the two wave-functions impinging from the two input gates of BS.  The superposition principle forces the resulting wave-packet to end its journey in the exit gate  $\gamma_+(t_f)$.  The pilot-wave dynamics is strongly impacted by the linearity of the superposition (compare with Fig.~\ref{Fig1}).} \label{Fig3}
\end{center}
\end{figure}
From the point of view of the BBQT (as illustrated in Fig. \ref{Fig3}) we have a mapping $\gamma_+(t_{in})\cup\gamma_-(t_{in})\rightarrow \gamma_+(t_f)$ which has nothing to do with either
Eq. \ref{mod1} and Eq. \ref{mod2} nor even Eq. \ref{mod3}. More precisely, the mapping associated with Eq. \ref{sum2} reads: 
\begin{eqnarray}
y(t_f)=\frac{y(t_{in})}{2}+\frac{1}{2}
\label{mod4}
\end{eqnarray} Therefore, the high contextuality of the BBQT leads (in agreement with wave-particle duality) to new features induced by the coherence of the different branches of the input wave function.\\
\indent In order to make sense of the Bernoulli shift Eq. \ref{mod3} in a simple way we modify the properties of our beam-splitter by adding phase plates in the input and output channels and while this is not mandatory for the generality of the reasoning we will from now consider instead of Eq \ref{arr} the unitary relation 
\begin{eqnarray}
  \left(\begin{array}{c}
  a'_+\\a'_-
  \end{array}\right)=\frac{1}{\sqrt{2}}\left(\begin{array}{cc}
  1 &1\\ 1 & -1
  \end{array}\right)\left(\begin{array}{c}
  a_+\\a_-
  \end{array}\right).\label{newBS}
 \end{eqnarray}
\begin{figure}[h]
\begin{center}
\includegraphics[width=10cm]{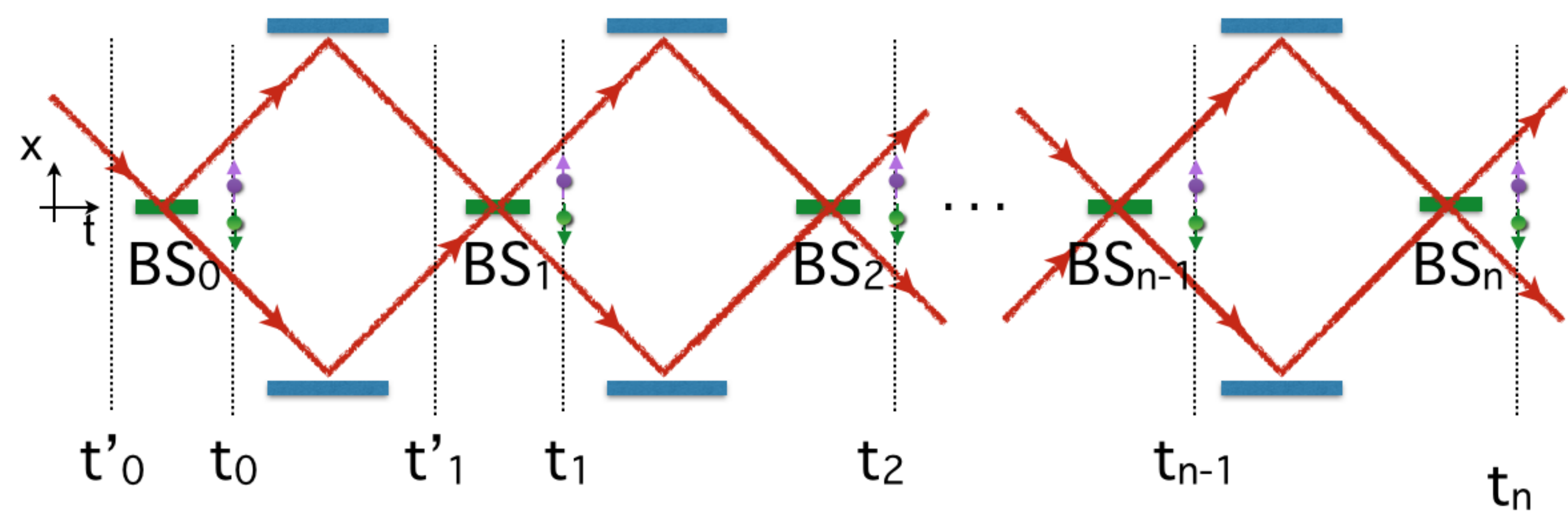} 
\caption{Skecth of the iterative procedure for entangling a initial wave-packet with `Bohmian' pointers  providing an unambiguous which- path information on the pilot wave particle motion (as explained in the main text). The various pointers interacting at time $t_0$, $t_1$ ... are sketched as qubit states.  } \label{Fig4}
\end{center}
\end{figure}
 Furthermore, in order the break the coherence between the two input waves $\psi_0(x,t_{in})$ and $\psi_1(x,t_{in})$ we introduce entanglement with an external pointer qubit before entering the beam splitter.  The pointer must represent an unambiguous  `which-path' information concerning the moving particle in the context of the BBQT.  We represent the initial state of the pointer by a wavefunction $\varphi^1_{in}(Z_1)$ associated with the coordinate $Z_1$ of the pointer (we assume $\int dZ_1 |\varphi^{1}_{in}(Z_1)|^2=1$). The interaction leading to entanglement works in the following way: Starting with an arbitrary state like $A\psi_0(x,t_0)+B\psi_1(x,t_0)$ at time $t_0$ and a fixed initial pointer state $\varphi^1_{in}(Z_1)$ we obtain:
  \begin{eqnarray}
 (A\psi_0(x,t_0)+B\psi_1(x,t_0)\varphi^1_{in}(Z_1)
 \rightarrow A\psi_0(x,t_0)\varphi^1_{\uparrow}(Z_1)+B\psi_1(x,t_0)\varphi^1_{\downarrow}(Z_1).\label{qubit}
 \end{eqnarray} Here we assume $\int dZ_1|\varphi^1_{\uparrow}(Z_1)|^2=\int dZ_1|\varphi^1_{\downarrow}(Z_1)|^2=1$ and $\int dZ_1\varphi^1_{\uparrow}(Z_1)(\varphi^1_{\downarrow}(Z_1))^\ast=0$.  Additionally, in order to simplify the analysis we suppose the pointer-particle interaction to be quasi-instantaneous and acting only at time $t\simeq t_0$. Moreover, in BBQT the positions of the particle and pointer play a fundamental ontic role. In order to have a genuine  Bohmian which-path information we thus require that the two pointer wavefunctions are well localized and are not overlapping, i.e., $\varphi^1_{\downarrow}(Z_1)\varphi^1_{\uparrow}(Z_1)=0$ $\forall Z_1$. \\
 \indent We now consider the following sequences of processes which are sketched on Fig.~\ref{Fig4}. First, we prepare  a non-entangled quantum system in the initial state $\psi_0(x,t'_0)\varphi_{in}(Z)$ with $t'_0\ll t_0$. Before interacting with the qubit the  particle wave-packet interacts with a first beam-splitter BS$_0$ like in the previous subsection. Using Eq.~\ref{newBS} and  Eq.~\ref{qubit} this leads to:
  \begin{eqnarray}
   \psi_0(x,t'_0)\varphi^1_{in}(Z_1)\rightarrow \frac{\psi_1(x,t_0)+\psi_0(x,t_0)}{\sqrt{2}}\varphi^1_{in}(Z_1)
   \rightarrow\frac{\psi_1(x,t_0)\varphi^1_{\uparrow}(Z_1)+\psi_0(x,t_0)\varphi^1_{\downarrow}(Z_1)}{\sqrt{2}}.
   \end{eqnarray} In order to use a probabilistic interpretation, i.e., Born's rule we impose the normalization $C=1/\sqrt{L}$ associated with the wave-packet $\Phi_0$ (see Eq.~\ref{num0}).  Second, as shown on Fig.~\ref{Fig4} the two wave-packets are moving in free space and interact with two mirrors which reflect the beams into the direction of a second beam-splitter BS$_1$ where they cross (BS$_1$ is the time translation of the same beam-splitter but we will continue to use this notation for simplicity). The main effect of the mirrors is to reverse the direction of propagation of $\psi_0(x,t_0)$ and $\psi_1(x,t_0)$, i.e., $\psi_0(x,t_0)\rightarrow -\psi_1(x,t'_1+\frac{2D}{v_x})e^{i\chi}$ and  $\psi_1(x,t_0)\rightarrow -\psi_0(x,t'_1+\frac{2D}{v_x})e^{i\chi}$ with $t'_1$ a time after the interaction and $\chi=\frac{2D}{v_x}(\omega_k-k_xv_x)$ a phase shift depending on the distance $D$ between BS$_0$ and any of the two mirrors ($-\frac{2D}{v_x}>0$ is the travel-time taken by the center of the wave-packet for moving from BS$_0$ to BS$_1$). At a time $t'_1$ before crossing BS$_1$ the quantum state reads thus  
   \begin{eqnarray}
   -e^{i\chi}\frac{\psi_0(x,t'_1+\frac{2D}{v_x})\varphi^1_{\uparrow}(Z_1)+\psi_1(x,t'_1+\frac{2D}{v_x})\varphi^1_{\downarrow}(Z_1)}{\sqrt{2}}.
   \end{eqnarray} 
 At a time $t_1\gg -\frac{2D+L}{v_x}$ after the interaction with BS$_1$ the quantum state reads (omitting the irrelevant phase factor)    
 \begin{eqnarray}
  \frac{\psi_0(x,t'_1+\frac{2D}{v_x})\varphi^1_{\uparrow}(Z_1)+\psi_1(x,t'_1+\frac{2D}{v_x})\varphi^1_{\downarrow}(Z_1)}{\sqrt{2}}
   \nonumber\\ \rightarrow \frac{\psi_1(x,t_1+\frac{2D}{v_x})\varphi^1_{\rightarrow}(Z_1)+\psi_0(x,t_1+\frac{2D}{v_x})\varphi^1_{\leftarrow}(Z_1)}{\sqrt{2}}
   \end{eqnarray}  
  where $\varphi^1_{\rightarrow}=\frac{\varphi^1_{\uparrow}+\varphi^1_{\downarrow}}{\sqrt{2}}$ and $\varphi^1_{\leftarrow}=\frac{\varphi^1_{\uparrow}-\varphi^1_{\downarrow}}{\sqrt{2}}$ are two orthogonal eigenstates. Now, if we write this quantum state during the interaction with $BS_1$ as  $\Psi(x,Z,t)=\psi_\uparrow(x,t)\varphi^1_{\uparrow}(Z_1)+\psi_\downarrow(x,t)\varphi^1_{\downarrow}(Z_1)$ we can define the Bohmian particle velocity $\frac{d}{dt}x^\Psi(t)=v(x,Z,t)$ as:
  \begin{eqnarray}
  \frac{d}{dt}x^\Psi(t)=\frac{v_\uparrow(x,t)|\psi_\uparrow(x,t)\varphi^1_{\uparrow}(Z_1)|^2+v_\downarrow(x,t)|\psi_\downarrow(x,t)\varphi^1_{\downarrow}(Z_1)|^2}{|\psi_\uparrow(x,t)\varphi^1_{\uparrow}(Z_1)|^2+|\psi_\downarrow(x,t)\varphi^1_{\downarrow}(Z_1)|^2}\label{velocityBS}
  \end{eqnarray} where we introduced the two velocities  $v_{\uparrow/\downarrow}(x,t)=\frac{1}{m}\partial_xS_{\uparrow/\downarrow}(x,t)$ associated with the two wavefunctions $\psi_{\uparrow/\downarrow}(x,t)$. Eq.~\ref{velocityBS} relies on the `which-path' constraint $\varphi^1_{\downarrow}(Z_1)\varphi^1_{\uparrow}(Z_1)=0$ and therefore we have here two different dynamics depending on the pointer position $Z_1$. If $Z_1$ lies in the support  of $\varphi^1_{\uparrow}(Z_1)$ we have the dynamics  $\frac{d}{dt}x^\Psi(t)=v_\uparrow(x,t)$ corresponding to Fig.~\ref{Fig1}(a) whereas if $Z_1$ lies in the support  of $\varphi^1_{\downarrow}(Z_1)$ we have the dynamics  $\frac{d}{dt}x^\Psi(t)=v_\downarrow(x,t)$ corresponding to Fig.~\ref{Fig1}(b).\\
\indent The previous procedure for generating decohered Bohmian paths can be repeated iteratively at the times $t_2$, $t_3$,...  after interaction with the beam-splitter BS$_2$, BS$_3$...(see Fig.~\ref{Fig4}). For this purpose   we consider at time $t_1$ entanglement with a an additional pointer initially in the state $\varphi^2_{in}(Z_)$ and we assume the  transformation:
\begin{eqnarray}
\frac{\psi_1(x,t_1+\frac{2D}{v_x})\varphi^1_{\rightarrow}(Z_1)+\psi_0(x,t_1+\frac{2D}{v_x})\varphi^1_{\leftarrow}(Z_1)}{\sqrt{2}}\varphi^2_{in}(Z_2)\nonumber\\
   \rightarrow\frac{\psi_1(x,t_1+\frac{2D}{v_x})\varphi^1_{\rightarrow}(Z_1)\varphi^2_{\uparrow}(Z_2)+\psi_0(x,t_1+\frac{2D}{v_x})\varphi^1_{\leftarrow}(Z_1)\varphi^2_{\downarrow}(Z_2)}{\sqrt{2}}.
   \end{eqnarray}  The wave-packets propagate into the interferometer and between the time ${t'}_2$ and $t_2$ we obtain 
   \begin{eqnarray}
\frac{\psi_0(x,t'_2+\frac{4D}{v_x})\varphi^1_{\rightarrow}(Z_1)\varphi^2_{\uparrow}(Z_2)+\psi_1(x,t'_2+\frac{4D}{v_x})\varphi^1_{\leftarrow}(Z_1)\varphi^2_{\downarrow}(Z_2)}{\sqrt{2}}\nonumber\\
   \rightarrow \frac{\psi_1(x,t_2+\frac{4D}{v_x})\varphi^{12}_{\rightarrow}(Z_1,Z_2)+\psi_0(x,t_2+\frac{4D}{v_x})\varphi^{12}_{\leftarrow}(Z_1,Z_2)}{\sqrt{2}}  \end{eqnarray} with the orthonormal  states $\varphi^{12}_{\rightarrow}=\frac{1}{\sqrt{2}}(\varphi^1_{\rightarrow}\varphi^2_{\uparrow}+\varphi^1_{\leftarrow}\varphi^2_{\downarrow})$ and $\varphi^{12}_{\leftarrow}(Z_1,Z_2)=\frac{1}{\sqrt{2}}(\varphi^1_{\rightarrow}\varphi^2_{\uparrow}-\varphi^1_{\leftarrow}\varphi^2_{\downarrow})$.\\ \indent This can be generalized at any time $t_n$ after  interaction with BS$_n$:
   \begin{eqnarray}
\frac{\psi_0(x,t'_n+\frac{2nD}{v_x})\varphi^{1,...,n-1}_{\rightarrow}(Z_1,...,Z_{n-1})\varphi^n_{\uparrow}(Z_n)+\psi_1(x,t'_n+\frac{2nD}{v_x})\varphi^{1,...,n-1}_{\leftarrow}(Z_1,...,Z_{n-1})\varphi^n_{\downarrow}(Z_n)}{\sqrt{2}}\nonumber\\
   \rightarrow \frac{\psi_1(x,t_n+\frac{2nD}{v_x})\varphi^{1,...,n}_{\rightarrow}(Z_1,...,Z_n)+\psi_0(x,t_n+\frac{2nD}{v_x})\varphi^{1,...,n}_{\leftarrow}(Z_1,...,Z_n)}{\sqrt{2}}, \nonumber\\ \label{term1}\end{eqnarray}
   with the orthonormal states $\varphi^{1,...,n}_{\rightarrow/\leftarrow}=\frac{1}{\sqrt{2}}(\varphi^{1,...,n-1}_{\rightarrow}\varphi^n_{\uparrow}\pm\varphi^{1,...,n-1}_{\leftarrow}\varphi^n_{\downarrow})$.
 Like for the interaction at BS$_1$ (between $t'_1$ and $t_1$) we can define a Bohmian dynamical evolution similar to  Eq.~\ref{velocityBS} but based on the wave function \begin{eqnarray}
 \Psi(x,Z1,...Z_n,t)=\psi_\uparrow(x,t)\varphi^{1,...,n-1}_{\rightarrow}(Z_1,...,Z_{n-1})\varphi^n_{\uparrow}(Z_n)\nonumber\\+\psi_\downarrow(x,t)\varphi^{1,...,n-1}_{\leftarrow}(Z_1,...,Z_{n-1})\varphi^2_{\downarrow}(Z_n).\label{term2}
 \end{eqnarray} We obtain the velocity  
 \begin{eqnarray}
  \frac{d}{dt}x^\Psi(t)=\frac{v_\uparrow(x,t)|\psi_\uparrow(x,t)\varphi^{1,...,n-1}_{\rightarrow}(Z_1,...,Z_{n-1})\varphi^n_{\uparrow}(Z_n)|^2}{|\psi_\uparrow(x,t)\varphi^{1,...,n-1}_{\rightarrow}(Z_1,...,Z_{n-1})\varphi^n_{\uparrow}(Z_n)|^2+|\psi_\downarrow(x,t)\varphi^{1,...,n-1}_{\leftarrow}(Z_1,...,Z_{n-1})\varphi^2_{\downarrow}(Z_n)|^2}\nonumber\\
  +\frac{v_\downarrow(x,t)|\psi_\downarrow(x,t)\varphi^{1,...,n-1}_{\leftarrow}(Z_1,...,Z_{n-1})\varphi^2_{\downarrow}(Z_n)|^2}{|\psi_\uparrow(x,t)\varphi^{1,...,n-1}_{\rightarrow}(Z_1,...,Z_{n-1})\varphi^n_{\uparrow}(Z_n)|^2+|\psi_\downarrow(x,t)\varphi^{1,...,n-1}_{\leftarrow}(Z_1,...,Z_{n-1})\varphi^2_{\downarrow}(Z_n)|^2}\nonumber\\ \label{velocityBSn}
  \end{eqnarray} which like Eq.~\ref{velocityBS} reduces to the one of the two dynamics i) $\frac{d}{dt}x^\Psi(t)=v_\uparrow(x,t)$ if $Z_n$ lies in the support  of $\varphi^n_{\uparrow}(Z_n)$ (i.e., corresponding to Fig.~\ref{Fig1}(a)), or ii) $\frac{d}{dt}x^\Psi(t)=v_\downarrow(x,t)$ if $Z_n$ lies in the support  of $\varphi^n_{\downarrow}(Z_n)$ (i.e., corresponding to Fig.~\ref{Fig1}(b)). The full history of the particle in the interferometer depends on the positions  $Z_1,...,Z_n$ taken by the various Bohmian pointers. In turn   this deterministic  iterative process allows us to define a Bernoulli map for the evolution.\\
\subsection{Mixing, chaos, and relaxation to quantum equilibrium} \label{section33}
\begin{figure}[h]
\begin{center}
\includegraphics[width=10cm]{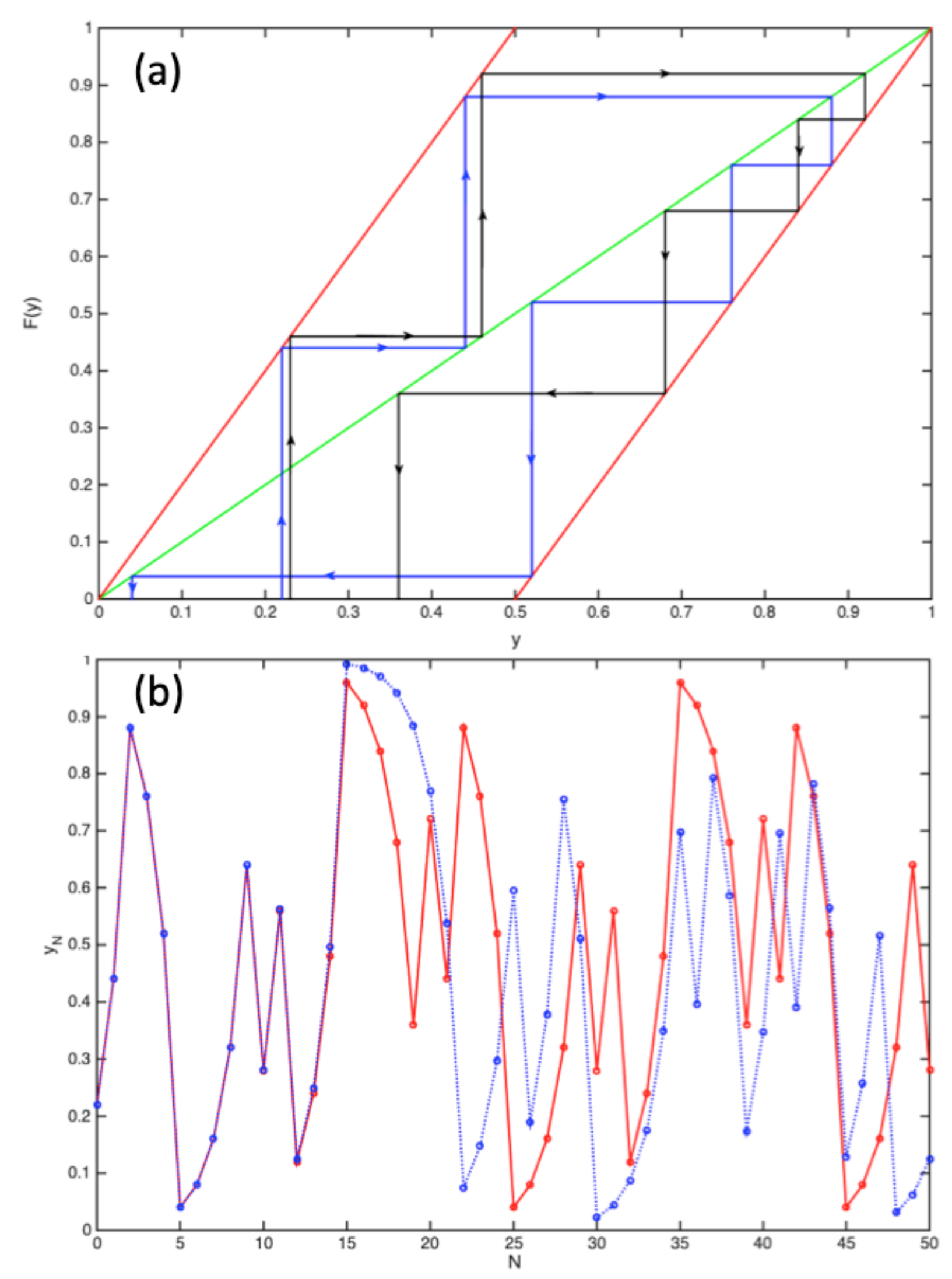} 
\caption{(a) Bernoulli map $y_n=F(y_{n-1})$ in the plane $y,y'$ where the function $y'=F(y)$ acts iteratively. The red and green line are acting as mirrors during the process. The black and blue trajectories correspond to different initial coordinates $y_0=0.22$ and $y_0=0.23$. (b) The same Bernoulli map is shown a as function $y=y(n)$ of the iteration steps $n=0,1,...$. The two chaotic trajectories shown in red and blue respectively correspond to $y_0=0.22$ and $y_0=0.220001$ (see main text). } \label{Fig5}
\end{center}
\end{figure}

\indent The Bernoulli map is clearly defined from Eqs.~\ref{mod3},\ref{mod3b} after introducing the variable $y(t)$ replacing  $x(t)$.
Between $t'_n$ and $t_n$ this reads:
\begin{eqnarray}
y(t_n)=2y(t'_n)&  \textrm{mod}(1). 
 \label{mod3ntemp}
 \end{eqnarray}  
  Moreover, the $y(t'_n)$ coordinate at time $t'_n$ is obviously equal to  $y(t_{n-1})$ at time $t_{n-1}$ (see Fig.~\ref{Fig4}) and therefore we have the map  
  \begin{eqnarray}
y(t_n)=2y(t_{n-1})&  \textrm{mod}(1). 
 \label{mod3n}
 \end{eqnarray} 
 This  iterative Bernoulli map $y_n=F(y_{n-1})$ is one of the simplest and most known chaotic map discussed in the literature~\cite{Driebe,Schuster}. In particular, its chaotic nature has been already studied in the context of the BBQT~\cite{Goldstein1992,Dewdney1996} (for different purposes as the one considered here) and an attempt to use it for deriving Born's rule has been worked out~\cite{Geiger} (without the entanglement used here and in \cite{Philbin2015,Drezet2017}).\\
 \indent The chaotic nature of the map is easy to obtain. Consider for example Fig.~\ref{Fig5}. 
In Fig.~\ref{Fig5}(a) we show a standard representation of the iterative function $y_n=F(y_{n-1})$ for two paths initially starting at $y_0=0.22$ and $y_0=0.23$ and after few iterations the coordinates are apparently diverging in a unpredictable way.  This is even more clear in the representation of Fig.~\ref{Fig5}(b) where two trajectories $y(t_n):=y_n$ are shown with  $y_0=0.22$ and $y_0=0.220001$. Again, the motions becomes chaotic after few iterations and the trajectories are strongly diverging. Mathematically, any number $y$ in the interval $[0,1]$ is represented in binary notations as $0.u_1u_2...u_n...$, i.e.,  $y=\frac{u_{1}}{2}+\frac{u_{2}}{4}+...+\frac{u_{n}}{2^n}+...$ where $u_n=0$ or 1. The Bernoulli transformation  $y'=F(y)$  with $y'=\frac{u'_{1}}{2}+\frac{u'_{2}}{4}+...+\frac{u'_{n}}{2^n}+...$ corresponds to the shift $u'_n=u_{n-1}$ i.e., to the binary number $0.u_2u_3...u_{n-1}...$. Iteratively this generates chaos since if the $n^{th}$ term in $y=\frac{u_{1}}{2}+\frac{u_{2}}{4}+...+\frac{u_{n}}{2^n}+...$ is known with an uncertainty $\delta y_i=\frac{1}{2^n}$ after $n$ iterations this uncertainty will grow up to $\delta y_f=1/2$. For example if $n=133$ and $\delta y_i=2^{-133}\simeq 10^{-40}$ we have after only 40 iterations completely lost any predictability in this dynamics (note that rational numbers are periodical in the binary representation and therefore the sequence will reappear periodically for rational numbers  representing a null measure in the segment $[0,1]$). It can be shown that this feature leads to randomness in close analogy with ideal probabilistic coin tossing~\cite{Ford}.  Therefore, any uncertainty will ultimately lead to chaos.  The Lyapunov divergence of this Bernoulli map  is readily obtained by considering as in Fig.~\ref{Fig5} two trajectories $y^{(A)}_n$ and $y^{(B)}_n=y^{(A)}_n+\delta y_n$ differing by a infinitesimal number such that 
\begin{eqnarray}
\delta y_n=2\delta y_{n-1}=2^n\delta y_0=e^{n\ln{2}}\delta y_0\label{Lyapounoff}
\end{eqnarray} where the positive Lyapunov exponent $\ln{2}$ characterizes this chaotic dynamics. If we introduce the time delay $\delta t =-2D/v_x>0$ and define the evolution time as $t_n=n\delta t$  we can rewrite the exponential divergence in Eq.~\ref{Lyapounoff} as $e^{+t/\tau}$ where $\tau=\frac{\delta t}{\ln{2}}$ defines a Lyapunov time.    \\
\indent Most importantly, the Bernoulli shift allows us to define a mixing property for the probability distribution $\rho(y)$. More precisely we can consider at any time $t_n$ the probability density $\rho(x,t_n)=\int...\int\rho(x,Z_1,...,Z_n,t_n)dZ_1...dZ_n$  where according to the BBQT we have $\rho(x,Z_1,...,Z_n,t_n)=f(x,Z_1,...,Z_n,t_n)|\Psi(x,Z_1,...Z_n,t_n)|^2$. In this framework  $\rho(x,t_n)$ is a coarse-grained probability involving a form of classical ignorance. In the following we will suppose that the pointers are all in quantum equilibrium and therefore we have $f(x,Z_1,...,Z_n,t_n):=f(x,t_n)$ and  $\rho(x,t_n)dx=f(x,t_n)d\Gamma(x,t_n)$ with  $d\Gamma(x,t_n)=dx\int...\int |\Psi(x,Z_1,...Z_n,t_n)|^2dZ_1...dZ_n$. \\
\indent For the present purpose a key result of deterministic maps like $y_n=F(y_{n-1})$ is the Perron-Frobenius theorem~\cite{Driebe,Schuster} which allows us to introduce the operator $\hat{U}_{PF}$, i.e., $\mu(y,t_{n+1})=\hat{U}_{PF}\mu(y,t_{n})$ with be definition $\rho(x,t)dx=\mu(y,t)dy$. For this we use the property for a trajectory
\begin{eqnarray}
\delta(y-y_{n+1})=\delta(y-F(y_n))=\int_0^1 dY\delta(y-F(Y))\delta(Y-y_n)\label{trucmuch}
\end{eqnarray} and the fact that any density $\mu(w,t_n)$ reads \begin{eqnarray}
\int_0^1dy(t_n)\mu(y(t_n),t_n)\delta(w-y(t_n))=\int_0^1dy(t_0)\mu(y(t_0),t_0)\delta(w-y(t_n))
\end{eqnarray} (where we used Liouville's theorem $dy(t_n)\mu(y(t_n),t_n)=dy(t_0)\mu(y(t_0),t_0)$).  Therefore, from Eq.~\ref{trucmuch} we obtain 
\begin{eqnarray}
\mu(y,t_{n+1})=\hat{U}_{PF}\mu(y,t_{n})=\int_0^1 dY\delta(y-F(Y))\mu(Y,t_n)\label{trucmuchB}
\end{eqnarray}
which for the Bernoulli map means 
\begin{eqnarray}
\mu(y,t_{n+1})=\hat{U}_{PF}\mu(y,t_{n})=\frac{1}{2}\left[\mu(\frac{y}{2},t_n)+\mu(\frac{y+1}{2},t_n)\right].\label{trucmuchC}
\end{eqnarray}
Moreover, for the present wavefunction defined in term of the wave-packet $\Phi_0(x)$ which is constant in amplitude in its support we can also write  
\begin{eqnarray}
\tilde{f}(y,t_{n+1})=\hat{U}_{PF}\tilde{f}(y,t_{n})=\frac{1}{2}\left[\tilde{f}(\frac{y}{2},t_n)+\tilde{f}(\frac{y+1}{2},t_n)\right]\label{trucmuchD}
\end{eqnarray} with by definition $f(x,t)=\tilde{f}(y,t)$ using the transformation $x\rightarrow y$ (see Eqs.~\ref{transf1},\ref{transf2}) and where $\int_{\gamma_+(t_n)\cup\gamma_-(t_n)} dx \frac{|C|^2}{2}f(x,t_n)=\int_0^1dy\tilde{f}(y,t_n)=1$ involving the normalization $C=1/\sqrt{L}$. This iterative Perron-Frobenius relation admits Bernoulli polynomial eigenstates defined by $\frac{1}{2^n}B_n(y)=\hat{U}_{PF}B_n(y)$ with $B_0(y)=1$, $B_1(y)=y-1/2$, $B_2(y)=y^2-y+1/6$,...\cite{Driebe}.\\
\indent It can be shown~\cite{Driebe} that the $B_m(y)$ polynomials form a basis for the probability function $\tilde{f}(y,t)$ and therefore it we we write $\tilde{f}(y,t_0)=\sum_{m=0}^{m=+\infty}A_mB_m(y)$ we have after $n$ iterations of the $\hat{U}_{PF}-$operator: 
\begin{eqnarray}
\tilde{f}(y,t_{n})=\sum_{m=0}^{m=+\infty}A_me^{-n\cdot m\ln{2}}B_m(y).\label{trucmuchE}
\end{eqnarray} In this formula  we have~\cite{Driebe,Prigogine2} 
\begin{eqnarray}
A_m=\int_0^1dy \tilde{f}(y,t_{0})\tilde{B}_m(y)\label{trucmuchF}
\end{eqnarray} where $\tilde{B}_0(y)=1$ and $\tilde{B}_m(y)=\lim_{\varepsilon\rightarrow 0^+}\frac{(-1)^{m-1}}{m!}\frac{d^{m-1}}{dy^{m-1}}[\delta(y-1+\varepsilon)-\delta(y-\varepsilon)]$ for $m\geq 1$.   This leads to $A_0=\int_0^1\tilde{f}(y,t_0)dy$ and $A_m=\lim_{\varepsilon\rightarrow 0^+}\frac{1}{m!}\frac{d^{m-1}}{dy^{m-1}}[\tilde{f}(1-\varepsilon,t_{0})-\tilde{f}(0+\varepsilon,t_{0})]$.
Eq.~\ref{trucmuchE} is important since it shows that in limit $n\rightarrow +\infty$ we have necessarily $\tilde{f}(y,t_{n})\rightarrow A_0B_0(y)=A_0$. Moreover, from the properties of the Bernoulli polynomials and the normalization of the probability density we have necessarily $\int_0^1\tilde{f}(y,t)dy=A_0=1$ (with $\int_0^1dy B_m(y)=\delta_{0,m}$).   Therefore we deduce
\begin{eqnarray}
\lim_{n\rightarrow+\infty}\tilde{f}(y,t_n)=\lim_{n\rightarrow+\infty}f(x,t_n)=1.\label{trucmuchG}
\end{eqnarray}  
\begin{figure}[h]
\begin{center}
\includegraphics[width=10cm]{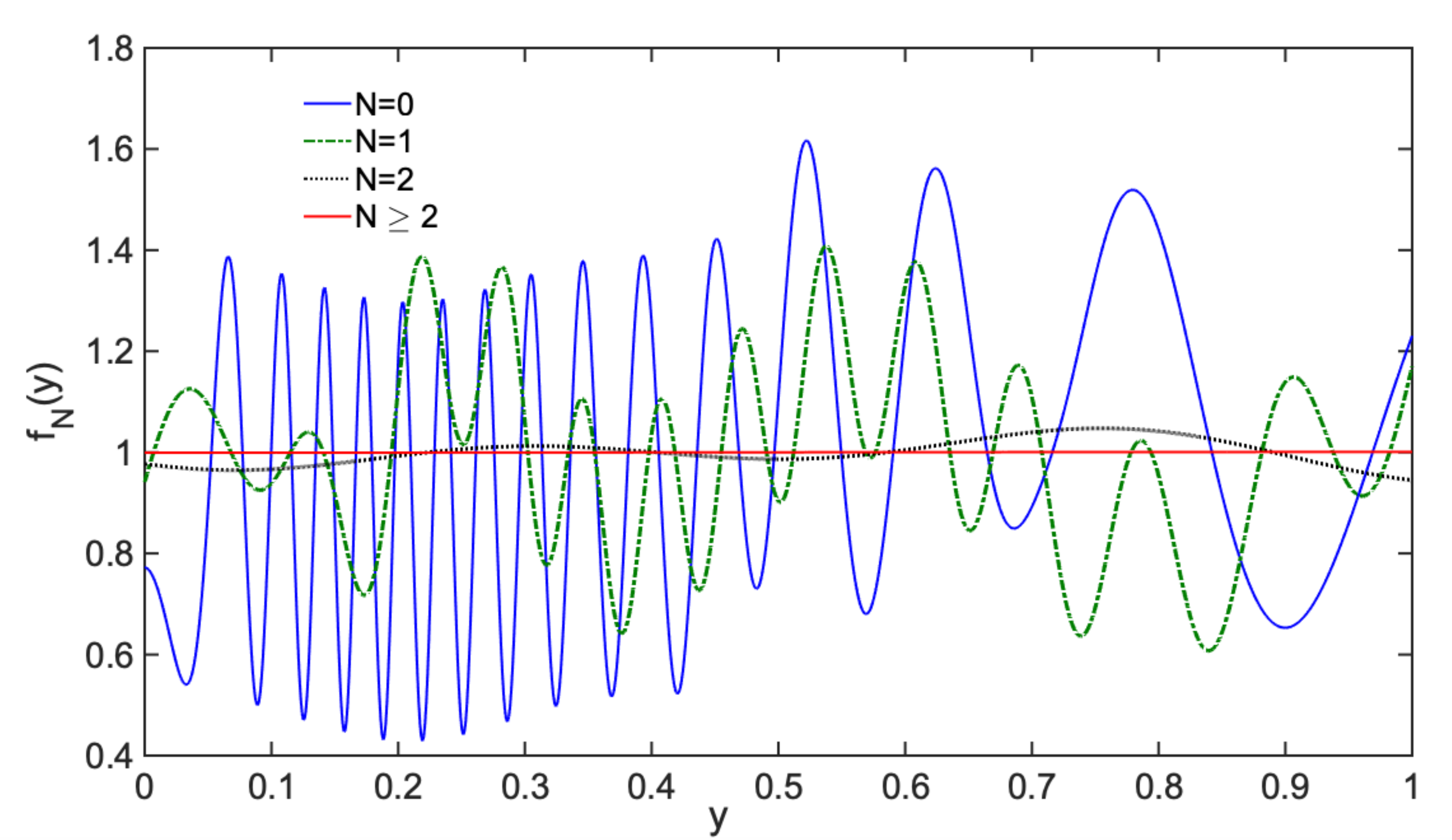} 
\caption{Evolution of $\tilde{f}(y,t_{n}):= \tilde{f}_n(y)$ as a function of $y$ for few $n$ values (using the Perron-Frobenius operator Eq.~\ref{trucmuchD}). The initial distribution $\tilde{f}_0(y)$ (blue curve)  was chosen to be arbitrarily  irregular. After few iterations $n\geq 2$ the function $\tilde{f}_n(y)$ can no be distinguished from the line $\tilde{f}=f=1$ associated with quantum equilibrium (i.e., Born's rule).   } \label{Fig6}
\end{center}
\end{figure}This result says that quantum equilibrium, and therefore Born's rule is a statistical attractor in the BBQT. Importantly, Eq.~\ref{trucmuchE} shows that each term in the sum is characterized by an exponential decay $e^{-mt_n/\tau}$ which is a signature of stability (negative Lyapunov exponent) whereas the trajectories (as we have shown in Eq.~\ref{Lyapounoff}) have a positive Lyapunov exponent associated with dynamical instability and chaos. These two pictures are thus clearly complementary. This was already emphasized long ago by Prigogine in a different context~\cite{Prigogine,Prigogine2}. As an illustration we show on Fig.~\ref{Fig6} the transformation of an arbitrary (normalized) density $\tilde{f}(y,t_0)$: after only three applications of the Perron-Frobenius operator the density  is indistinguishable from the quantum equilibrium $\tilde{f}=f=1$ which is therefore a very efficient attractor.\\
\indent We emphasize that the iterative process sketched in Fig.~\ref{Fig4} and associated with states like Eqs.~\ref{term1},\ref{term2} involves in the end two branches $\psi_0(x,t_n)$ and $\psi_1(x,t_n)$ entangled  with an environment of Bohmian pointers characterized by $\varphi^{1,...,n}_{\rightarrow/\leftarrow}=\frac{1}{\sqrt{2}}(\varphi^{1,...,n-1}_{\rightarrow}\varphi^n_{\uparrow}\pm\varphi^{1,...,n-1}_{\leftarrow}\varphi^n_{\downarrow})$. Moreover,  because of the orthogonality of theses pointer states the two branches $\psi_0(x,t_n)$ and $\psi_1(x,t_n)$ can not interfere: they are decohered.  Still, in each of the two final  wave-packets $\psi_0(x,t_N)$ and $\psi_1(x,t_N)$ (after a large number of iterations $N$) we have $f(x,t_N)\simeq 1$ with a high accuracy.   Therefore, supposing that we now make a pinhole for selecting one of these two branches   we have prepared a quantum system satisfying Born's rule  $\rho(x,t)\simeq |\psi(x,t)|^2$. Fundamentally, it means that if an entangled system like the one we discussed is postselected by a filtering procedure we can define subsystems for which Born rules is true and where quantum coherence is maintained (this is the case with our two wave functions $\psi_1$ and $\psi_0$ taken separately. For example the wave function $\psi_0(x,t)$ can be collimated and sent into an interferometer in order to observe wave-particle duality. All the systems following this guiding wave belong to a statistical ensemble of particles obeying Born's rule $f\simeq 1$. Therefore, all the predictions of standard quantum mechanics will be reproduced with these systems.  \\     
\indent Although the present model is rudimentary it allows us to obtain precious information on relaxation to quantum equilibrium.  Indeed, observe that  in the continuous time approximation we have  $\tilde{f}(y,t)\simeq 1+ A_1e^{-t/\tau}B_1(y)$ which is a solution of the differential equation 
\begin{eqnarray}
\frac{\partial \tilde{f}(y,t)}{\partial t}=-\frac{\tilde{f}(y,t)-1}{\tau}
\end{eqnarray} 
This suggests a collision term in a Boltzmann-like equation and therefore an extension of our model by writing
\begin{eqnarray}
\frac{df(x,t)}{dt}:=\partial_t f(x,t)+v_\psi(x,t)\partial_xf(x,t)=-\frac{f(x,t)-1}{\tau}
\end{eqnarray} or equivalently with $\rho(x,t)=f(x,t)|\psi(x,t)|^2$ and $\partial_t|\psi(x,t)|^2+\partial_x(v_\psi(x,t)|\psi(x,t)|^2)=0$:
\begin{eqnarray}
\partial_t \rho(x,t)+\partial_x(v_\psi(x,t)\rho(x,t))=-\frac{\rho(x,t)-|\psi(x,t)|^2}{\tau}.
\end{eqnarray}
 With such dynamics (with an effective broken time symmetry) it is useful to introduce the Valentini entropy~\cite{Valentini1991}: \begin{eqnarray}
  S_t:=-\int f(x,t)\ln{(f(x,t))}d\Gamma(x,t) 
\end{eqnarray}  with $d\Gamma(x,t)=|\psi(x,t)|^2$. From the previous equation we deduce
 \begin{eqnarray} 
\frac{d}{dt}S_t=-\int\frac{df_t}{dt}(1+\ln{f_t})d\Gamma_t=\int\frac{(f_t-1)}{\tau}(1+\ln{f_t})d\Gamma_t=\int\frac{(f_t-1)}{\tau}\ln{f_t}d\Gamma_t.
\end{eqnarray}
Such a kinetic equation leads to a quantum version of the Boltzmann H-theorem as it can be shown easily: First, we have by definition $a\ln{b}+\frac{a}{b}-a\geq 0$ (with $a,b>0$) leading to  $(f-1)\ln{f}+\frac{f-1}{f}-f-1\geq 0$ if $f-1>0$, i.e., we obtain $(f-1)\ln{f}\geq \frac{(f-1)^2}{f}$ if $f-1>0$. Moreover, we have also $\ln{f}\leq f-1$ and thus if $f-1<0$ we have $(f-1)\ln{f}\geq (f-1)^2$. Now, separating the full $\Gamma-$ space into two parts $\Gamma_+$ and $\Gamma_-$ where $f-1\geq 0$ and $1-f\geq 0$ respectively we have: 
 \begin{eqnarray} 
\frac{d}{dt}S_t=\int\frac{(f_t-1)}{\tau}\ln{f_t}d\Gamma_t\geq \int_{\Gamma_+}\frac{(f_t-1)^2}{f_t\tau}d\Gamma_t+\int_{\Gamma_-}\frac{(f_t-1)^2}{\tau}d\Gamma_t\geq 0.
\end{eqnarray} Therefore, Valentini's entropy $S_t$ can not decrease  and the equality $\frac{d}{dt}S_t=0$ occurs iff $f=1$ corresponding to the quantum equilibrium. This defines a H-theorem for the BBQT.
\section{Conclusion and perspectives} \label{section4}
\indent The proposal discussed in this work is certainly schematic but it leads to several interesting conclusions.   First, since the dynamics maps used here is deterministic and chaotic it shows that randomness is unavoidable in the BBQT. As it was stressed by Prigogine~\cite{Prigogine,Prigogine2} we have here two complementary descriptions: one with trajectories which can be associated with the evolution map  $y_{n+1}=F(y_n)$ and the second with probability density i.e., as given by the Perron-Frobenius transformation $\tilde{f}(y,t_{n+1})=\hat{U}_{PF}\tilde{f}(y,t_{n})$.  The two pictures are  of course not independent since for a single trajectory  we have  $\delta(y-y_{n+1})=\hat{U}_{PF}\delta(y-y_{n})$ (i.e., $\tilde{f}(y,t_n)=\delta(y-y_{n})=f(x,t_n)=\frac{2}{|C|^2}\delta(x-x_n)=2L\delta(x-x_n)$). Moreover, for a trajectory the probability distribution is singular and the convergence to equilibrium is infinitely slow (this is connected to the fact that the coefficients $A_m$ in Eq.~\ref{trucmuchF} is given by an integral which is badly defined for the singular Dirac distribution $\tilde{f}(y,t_0)=\delta(y-y_{0})$). Therefore, the infinite precision required for computing such a chaotic path (due to the exponentially growing deviation errors with time) leads for all practical computations to the strong randomness above mentioned. To quote Ford~\cite{Ford} ``a chaotic orbit is random and incalculable; its information content is both infinite and incompressible''.  Subsequently, because of the extreme sensitivity in the initial conditions associated with the predictability horizon and the positive Lyapunov exponent,  the use of probability distributions in the BBQT  seems  (at least in our model) unavoidable if we follow Prigogine reasoning. Indeed, for Prigogine dynamic instability (and thus deterministic chaos) leads to probability. The necessarily finite precision $\delta y_0$ used for determining the position of a particle will grow exponentially with time to ultimately cover the whole segment $[0,1]$. Therefore, if we assign a uniform ignorance probability $\tilde{f}_0$ over the segment $\delta y_0$ (in which the particle is located) then, i.e., subsequently  after few iteration, we will have $\tilde{f}_t=1$ over the whole segment.   Yet, we stress that we don't here share all the conclusions obtained by Prigogine concerning determinism and probability (for related and much more detailed criticisms see e.g., \cite{Bricmont}). Indeed, the BBQT (like classical mechanics that was considered by Prigogine in ~\cite{Prigogine,Prigogine2}) is a fully deterministic theory with  a clear ontology in the 3D and configuration space. Therefore, while a trajectory could be incalculable by any finite mean or algorithm, the path still fundamentally exists for an idealized  Laplacian daemon, i.e., having access to an infinite computing power and precision for locating and defining the particle motion. This metaphor is the core idea of Einstein's realism:   Postulating the existence of a World independently of the presence or absence of observers (even if the observers can be part of the World).  From this ontic perspective  we need more than just ignorance in order to justify the use of probability in statistical physics.   Indeed, as emphasized long ago by Poincar\'e the laws of the kinetic theory of  gases still hold true even if we exactly know the positions of all molecules -~\cite{Poincare}. There is something of objective in the laws  of statistical mechanics that goes beyond mere ignorance: Otherwise how parameters such as diffusion constants would have objective physical contents? This point was emphasized by Prigogine from the very beginning and this constitutes the motivation for his  program in order to justify the objectivity of thermodynamics in general  and  the second law, i.e., irreversibility in particular. However, in our opinion the missing point in Prigogine implication: ``instability $\rightarrow$ probability $\rightarrow$ irreversibility'' is the recognition that in a deterministic theory the laws (chaotic or not) are not all but must be supplemented by specific initial  conditions  ultimately having a cosmological origin. Indeed, if we suppose a Universe made of only one electron  described initially by the wavefunction $\psi_0(x,t)$ and all the pointers involved in the iterative procedure sketched in Fig.~\ref{Fig4} then  we must use the chaotic Bernoulli map $y_{n+1}=F(y_n)$ for this system or equivalently the  Perron-Frobenius evolution $\delta(y-y_{n+1})=\hat{U}_{PF}\delta(y-y_{n})$. As we explained this system is unstable due to the presence of a positive Lyapunov exponent. Moreover, if we want to make sense of the formulas \ref{trucmuchD} and \ref{trucmuchE} with the rapid convergence  to $\tilde{f}=f=1$ we must consider a sufficiently regular distribution $\tilde{f}(y,t_0)\neq\delta(y-y_{0})$.  Now, as reminded in Sec.\ref{section2} the application of the WLLN to a statistical ensemble requires a `metric' of typicality associated with the Laplacian definition of probability.  In BBQT this metric reads $\rho(\mathbf{r},t)=f(\mathbf{r},t)|\psi(\mathbf{r},t)|^2$ and the law of large numbers leads  to Eq.~\ref{eq5b}, i.e.,   $\rho(\mathbf{r},t)\equiv \lim_{N \to +\infty}\frac{1}{N}\sum_{k=1}^{k=N}\delta^3(\mathbf{r}-\mathbf{x}^{\psi}_k(t))$  defined probabilistically in the long run, i.e, for an infinitely long sequence or infinite system. In our problem, this means that we consider an infinite Gibbs ensemble of copies similar to our system described in Fig.~\ref{Fig4}.   Here, the presence of an infinite sum of Dirac distributions is expected to lead to difficulties in connections with the chaotic map $\delta(y-y_{n+1})=\hat{U}_{PF}\delta(y-y_{n})$.   Therefore, in our problem if  the WLLN  $\rho(\mathbf{r},t)\equiv \lim_{N \to +\infty}\frac{1}{N}\sum_{k=1}^{k=N}\delta^3(\mathbf{r}-\mathbf{x}^{\psi}_k(t))$  is used to specify the initial distribution at time $t_0$ we must be aware that this should preserve the chaotic description  associated with the positive Lyapunov exponent and Dirac distributions should be problematic. In order to remove this unpleasant feature one must introduce coarse-graining as proposed by Valentini~\cite{Valentini1991,Valentini2020}. In our case this can be done by using a regular weighting function $\Delta(u)$ such that $\overline{\rho}(x,t)=\int du\delta(u)\rho(x-u,t)$ which in connection with the WLLN leads to $\overline{\rho}(x,t)\equiv \lim_{N \to +\infty}\frac{1}{N}\sum_{k=1}^{k=N}\Delta(x-x^{\psi}_k(t))$. Coarse-graining of cells in the configuration space plays a central role in the  work of Valentini for defining a ``subquantum H-theorem'' \cite{Valentini1991,ValentiniPhD}. Here, we see that in connection with Prigogine work coarse-graining must be supplemented with a dose of deterministic chaos and entanglement in order to reach the quantum equilibrium regime.  We believe that these two picture complete each other very well.  \\ 
\indent To summarize, in this work, we have proposed a mechanism for relaxation to quantum equilibrium in order to recover Born's rule in the BBQT. The proposed mechanism relies on entanglement with an environment of `Bohmian pointers' allowing the system to be mixing. The scenario was developed for the case of a single particle  in 1D motion interacting with beam splitters and mirrors but the model could be generalized to several situations involving collisions between quanta and scattering with defects or other particles. The general proposal is thus to consider the quantum relaxation to Born's rule as a genuine process in  phases of matter where interactions between particles play a fundamental role.  This involves usual condensed matters or even plasma or gases where collisions are mandatory. For example, if based on our toy model we consider that interaction with the beam splitter and entanglement with Bohmian pointers is a good qualitative model for discussing collisions between molecules in the atmosphere, and if we remember that nitrogen molecules at a temperature of 293 K and at a pressure of 1 bar involve typically a collision frequency of 7 $10^9$ /s (which implies a fast dynamics for reaching quantum equilibrium) we have thus a huge number  $n$ of collisions per second corresponding to a huge number of iterations in our Bernoulli-like process based on the Perron-Frobenius operator $f(y,t_{n+1})=\hat{U}_{PF}f(y,t_{n})$.  Compared  to Valentini's framework~\cite{Valentini1991,ValentiniPhD} where mixing and relaxation to quantum equilibrium are associated with coarse-graining \`a la Gibbs  our approach emphasizes the role  of information losses due to entanglement with a local environment.  In both cases,   we obtain an increase of entropy and a formulation of the $H-$theorem for the BBQT. Theses two views are certainly complementary in the same way that Gibbs and Boltzmann perspectives on entropy are related. This could have an impact on the efficiency of quantum relaxation in the early stages of the Universe evolution~\cite{Valentini2007,Valentini2015}. 
                 

\end{document}